\documentclass[conference]{IEEEtran}
\IEEEoverridecommandlockouts
\usepackage{cite}
\usepackage{amsmath,amssymb,amsfonts}
\usepackage{graphicx}
\usepackage{textcomp}
\usepackage{xcolor}
\usepackage{amsthm}

\usepackage{microtype}
\usepackage{algorithm}
\usepackage[noend]{algpseudocode}
\usepackage{comment}

\newtheorem{def5}{Note}

\def\BibTeX{{\rm B\kern-.05em{\sc i\kern-.025em b}\kern-.08em
    T\kern-.1667em\lower.7ex\hbox{E}\kern-.125emX}}
\begin{document}

\title{G-IOTA: Fair and confidence aware tangle\\
}

\author{\IEEEauthorblockN{Gewu Bu}
\IEEEauthorblockA{\textit{Sorbonne University, LIP6, UMR 7606} \\
Paris, France\\
gewu.bu@lip6.fr}
\and
\IEEEauthorblockN{\"{O}nder G\"{u}rcan}
\IEEEauthorblockA{\textit{CEA LIST} \\
Point Courrier 174 \\ 91191 Gif-sur-Yvette, France\\
onder.gurcan@cea.fr}
\and
\IEEEauthorblockN{Maria Potop-Butucaru}
\IEEEauthorblockA{\textit{Sorbonne University, LIP6, UMR 7606} \\
Paris, France \\
maria.potop-butucaru@lip6.fr}
}


\maketitle

\begin{abstract}
This paper proposes strategies to improve the IOTA tangle in terms of resilience to splitting attacks.
Our contribution is two fold. First,  we define  the notion of \emph{confidence fairness} for tips selection algorithms to guarantee the first approval for all honest tips. Then, we analyze  IOTA-tangle from the point of view of confidence fairness and identify its drawbacks.
Second, we propose  a new selection mechanism, G-IOTA,  that targets to protect tips  left behind. G-IOTA therefore has a good confidence fairness. G-IOTA lets honest transactions increase their confidence efficiently. 
Furthermore, G-IOTA includes an incentive mechanism for users who respect the algorithm and punishes conflicting transactions.
Additionally, G-IOTA provides  a mutual supervision mechanism that reduces the benefits of speculative and lazy behaviours. 
\end{abstract}

\begin{IEEEkeywords}
 IOTA - Tangle, Transaction confidence, Tips selection algorithm, Confidence fairness
\end{IEEEkeywords}

\section{Introduction}
Blockchain and distributed ledger technologies have emerged as one of the most revolutionary developments in recent years, with the goal of eliminating centralised intermediaries and installing distributed trusted services. They facilitate trustworthy trades and exchanges over the Internet, power cryptocurrencies, ensure transparency for documents, and much more. 
Traditionally, blockchain systems maintain a continuously-growing list of ordered blocks that include  one or more transactions that have been verified by the members of the system, called miners.  Blocks are linked using cryptography and  the order of blocks in the blockchain is the result of a form of agreement among the system participants. 

After the releasing of the most popular blockchains  (e.g., Bitcoin \cite{bitcoin} or Ethereum \cite{Ethereum}) with a specific focus on economical transactions their huge potential  for various other applications became evident.  
However, quickly after their release blockchains reached their limits in terms of throughput, blocksize and unforeseen behaviors. Therefore, 
non academic research further concentrate in  proposing alternatives by improving some  performance aspects but with non zero costs either on security or throughput. 
One of these solutions extends the blockchain  to a DAG overlay, provide a total ordering over all blocks and transactions, and outputs a consistent set
of accepted transactions.  In the research along this line (e.g., \cite{SompolinskyZ18,SompolinskyLZ16}) transactions are still organized on blocks. All these DAG-based systems structure blocks in a Directed Acyclic Graph.
Each block can include several references to predecessors.  

More recently, IOTA has been defined as an alternative dedicated to IoT area where  micro-transactions  are submitted at a very high frequence. 
 Transactions define a DAG overlay \emph{a.k.a.} tangle.  Strategies to maintain an IOTA-tagle have been proposed and analyzed in \cite{popov2016tangle}. IOTA-tangle properties have been formalized and analyzed in  \cite{kusmierz2017first}, \cite{popov2017equilibria} and \cite{kusmierz2018extracting}. More recently, \cite{staupequasi} and \cite{bramas:hal-01782959} analyze in detail the stability and the attack resilience  of IOTA-tangle. 
 One of the first works addressing the fairness of the selection mechanism in the IOTA tangle is \cite{kusmierz2018probability}. 

In this paper we introduce a new notion of fairness: \emph{confidence fairness} for tips selection algorithms to guarantee the first approval for all honest tips. We propose a new tangle, G-IOTA, that aims at increasing the overall fairness in IOTA-tangle by protecting tips who have been left behind, incentivizes users to respect the algorithm and punishes the conflicting transactions. G-IOTA provides also a mutual supervision mechanism that reduces the benefits of speculative and lazy behaviours. Moreover, G-IOTA tangle is as resistant as  IOTA-tangle to attacks, especially to splitting attacks.


The organization of this paper is as follows. 
Section \ref{sec:IOTA} introduces  IOTA and identifies its drawbacks.
Section \ref{sec:G-IOTA} proposes G-IOTA that is designed to overcome the drawbacks of IOTA-tangle. 
Section \ref{sec:Discussion} discusses the strengths and the weaknesses of G-IOTA.
Finally, Section \ref{sec:Conclusion-and-Future-Work} concludes the paper and discuss  future research directions.

\section{Background on IOTA}
\label{sec:IOTA}
In this section, we introduce the fundamental concepts of IOTA system (Sections \ref{sec:Tangle}, \ref{sec:Tip-Selection-Algorithm} and \ref{sec:Confidence-of-Transactions}). Furthermore,  we discuss  the drawbacks of its design  (Section \ref{sec:Drawbacks-of-IOTA}).

\subsection{IOTA Tangle}
\label{sec:Tangle}
IOTA is a permissionless distributed ledger that 
stores all the transactions going though the system, shared by all the \textit{users} in the IOTA network.
It utilizes a data structure called the \textit{tangle} \cite{popov2016tangle}.

The \textit{users} in IOTA work cooperatively to verify the transactions in the tangle for making the IOTA-tangle safe and reliable.
Users are grouped in two distinct subsets: \textit{honest users} that follow the IOTA protocol exactly, and \textit{malicious users} that show arbitrarily behaviors. 

\begin{figure}
\centering
\includegraphics[width=0.35\textwidth]{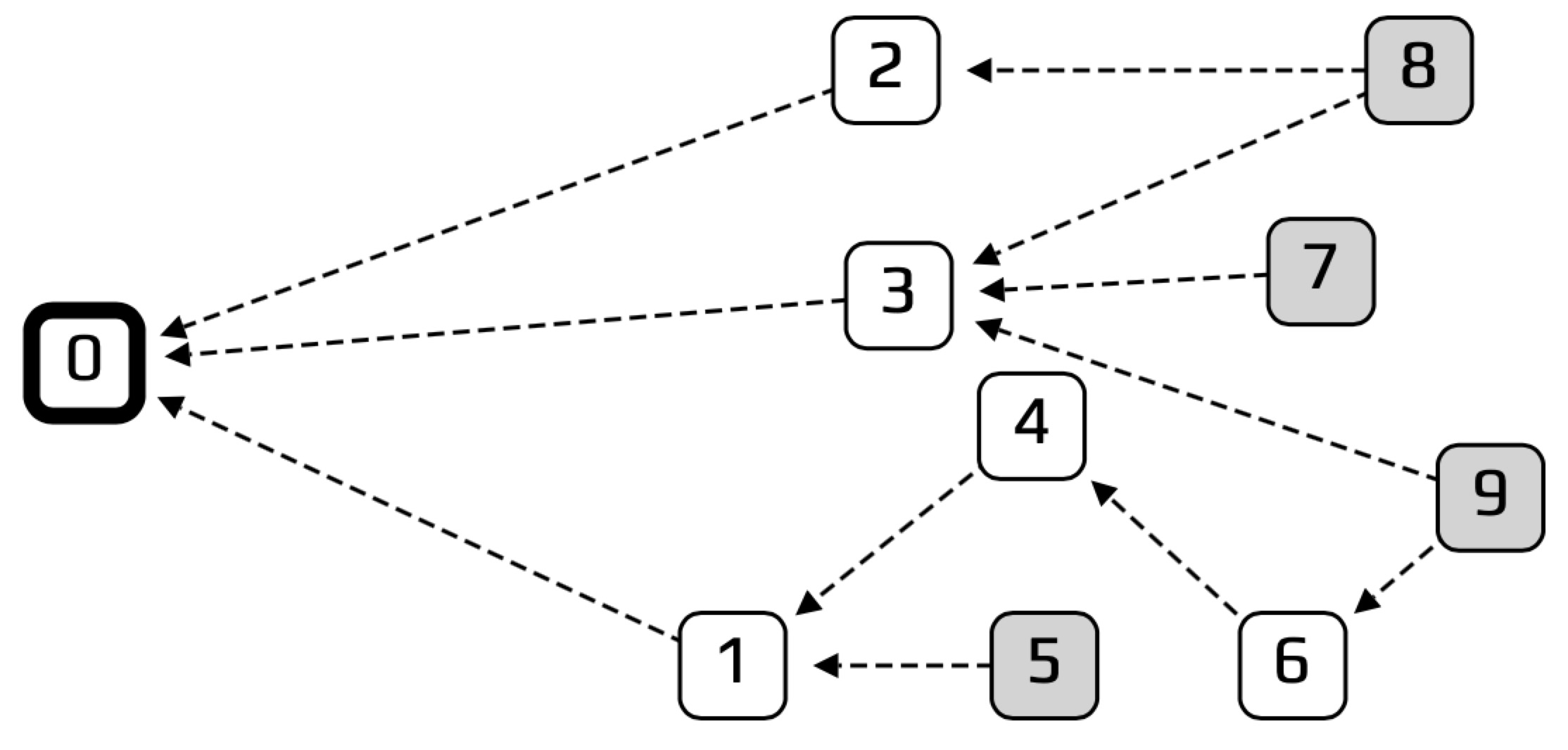}
\caption{An example of IOTA tangle}
\label{tangle}
\end{figure}

The \textit{tangle} is a Directed Acyclic Graph (DAG) where each vertex is a  transaction submitted to the IOTA network and the directed links represent a verification of transactions.
Figure \ref{tangle} shows an example IOTA tangle in which the white squared transactions are called \textit{verified transactions} and the grey squared ones are called \textit{unverified transactions} or \textit{tips}\footnote{The term \textit{tips} will be used hereafter throughout the paper.}.
The root of the tangle (transaction 0 in Figure \ref{tangle}) is called the genesis transaction.
The \textit{depth} $d_t$ of a transaction $t$ in the tangle is then defined as the length of the longest path from the genesis to the transaction. 
The \textit{depth} $D$ of the tangle, is then defined as $max(d_t) \quad \forall \quad t$, i.e. the maximum depth over all the transactions in the tangle. 

Users \textit{submit} transactions to the IOTA network.
The transactions submitted by honest users are called \textit{honest transactions}, and the transactions submitted by malicious users are called \textit{malicious transactions}.
A submitted transaction received by a user is called \textit{incoming transaction}. Since incoming transactions are not verified yet, they are also called \textit{incoming tips}. 

In all cases, to submit a new transaction to the tangle, two transactions from the tangle must be chosen for \textit{direct verification}. 
In general, the tips should be chosen for this purpose.
\textit{Indirect verification} means, on the other hand, verifying all the transactions in the \textit{verification path} of a tip/transaction. 
Formally, a \textit{verification path} for a transaction $t$ is a set of transactions $S_t$, containing all the transactions from $t$ to the genesis transaction. 
In Figure \ref{tangle}, the verification path of transaction $8$ is from $8$ to $2$ and $3$ and to $0$.
Furthermore, if a tip $t$ has not received its first verification yet and $D - d_t > d_S$ where $d_S$ is a configurable threshold, we say that the tip $t$ is \textit{left behind}.

It should be noted that unlike in the traditional blockchain systems, the verification of transactions is the responsibility of all users that want to submit transactions to the IOTA network. 
In other words, to submit transactions, each user must contribute to the IOTA network by verifying the previously submitted transactions.  Note that if there are conflicting transactions between two possible chosen tips, the sender needs to chose only one of them and chose another candidate tip to verify.

After verifying two tips, the sender of the transaction will broadcast its transaction and the chosen tips to all the other users in the network. By receiving submitted transactions from other users, the receivers update their local tangle by attaching the new transaction to the corresponding tips. 


\subsection{Tip selection algorithm}
\label{sec:Tip-Selection-Algorithm}

The method for choosing two tips for verifying is called the \textit{tip selection algorithm} and is an important decision for users. To resist to various types  of attacks and to reduce the negative impact of lazy nodes (see \cite{popov2016tangle}), three kind of \emph{tip selection algorithm} are proposed in \cite{popov2016tangle}: \emph{Uniform Random}, \emph{Unweighted Random Walk} and \emph{Weighted Random Walk} algorithms. From the analysis proposed in \cite{popov2016tangle}  the most advanced tip selection algorithm is weighted random walk algorithm. 

The \emph{Weighted Random Walk} is an application of Markov Chain Monte Carlo (MCMC) algorithms. 
The main idea is that multi-random walks are launched in the middle of the tangle. 
A random walk will go towards candidate tips following opposite direction of the arrows in the DAG (e.g., the opposite direction of the arrows in Figure \ref{tangle}). 
As a transaction in a tangle could have several directly verifying transactions, the cumulative weight of a transaction $v$ in a tangle is the total number of the transactions verifying $v$ directly or indirectly. 
The more cumulative weight a next transaction has, the higher probability the random walker will go to it. 

Weighted Random Walk has a configurable parameter $\alpha$ to control the effectiveness of the cumulative weight. With a high enough $\alpha$, even if the cumulative weights of two transactions has a small difference, the transaction with a bigger cumulative weight will have a higher probability to be chosen by the random walk. On the other hand, if $\alpha$ is small enough, even if the cumulative weights of two transactions have a big difference, they have almost the same probability to be chosen by the random walk.
 

The steps of the \emph{Weighted Random Walk} algorithm are as follows:
 
\begin{itemize}
   \item[1] Chose an interval in tangle, $[W, 2W]$, where $W$ represent the depth or the length of tangle.
   \item[2] Place $N$ random walks particles on transactions in that interval.
   \item[3] Let $N$ particles perform discrete direct weighted random walkers toward tips.
   \item[4] Chose the two tips reached first by the random walks.
\end{itemize}

Note that to avoid choosing \emph{lazy tips} that verify old transactions, IOTA may  discard  random walks who reach tips "too fast".

\subsection{Confidence of transactions}
\label{sec:Confidence-of-Transactions}

Transactions in tangle need to get enough (not only in terms of quantity) verifications to be finally considered as \textit{confirmed}. In Bitcoin-like blockchains, we say that a transaction $t$ can be confirmed with high probability, if there are at least $6$ blocks appended to the block that contain $t$.

The confidence of a transaction is a percentage between $0\%$ and $100\%$. A transaction in the tangle can be considered as confirmed, if it has has enough confidence. Note that "enough" depends on the relation between the participants to the transaction. For a transaction between friends, 80\% of confidence could be "enough" instead of 100\% for transaction between two strangers.

The \textit{confidence} of transaction is in relation to the tips selection algorithm. 
The general idea is that if a transaction has a high probability to be verified by new tips in the future, then this transaction will have a high confidence. 
For a given instant, $t$, we launch the tips selections algorithm $n$ times. Each run selects a tip. Let $i$ be one of the tips in the tangle at instant $t$, and $n_i$ is the number of time $i$ has been selected after launching $n$ times the tips selection algorithm. 
Given a transaction $s$, let $A_{s}$ contains all tips that verify $s$ directly or indirectly. 
The confidence of the transaction $s$, $C_s$ is:
\begin{align} 
\label{fff1}
C_s = \sum {n_i}/N \quad \forall i \in A_{s}
\end{align}


\subsection{Drawbacks of IOTA-tangle}
\label{sec:Drawbacks-of-IOTA}

In \cite{popov2016tangle} it is recommended that the Weighted Random Walk algorithm needs to chose a $\alpha$ relatively high in order to avoid  \emph{Splitting attack}.

\subsubsection{Splitting Attack}
\label{sec:Splitting-Attack}

\begin{figure}
\centering
\includegraphics[width=0.4\textwidth]{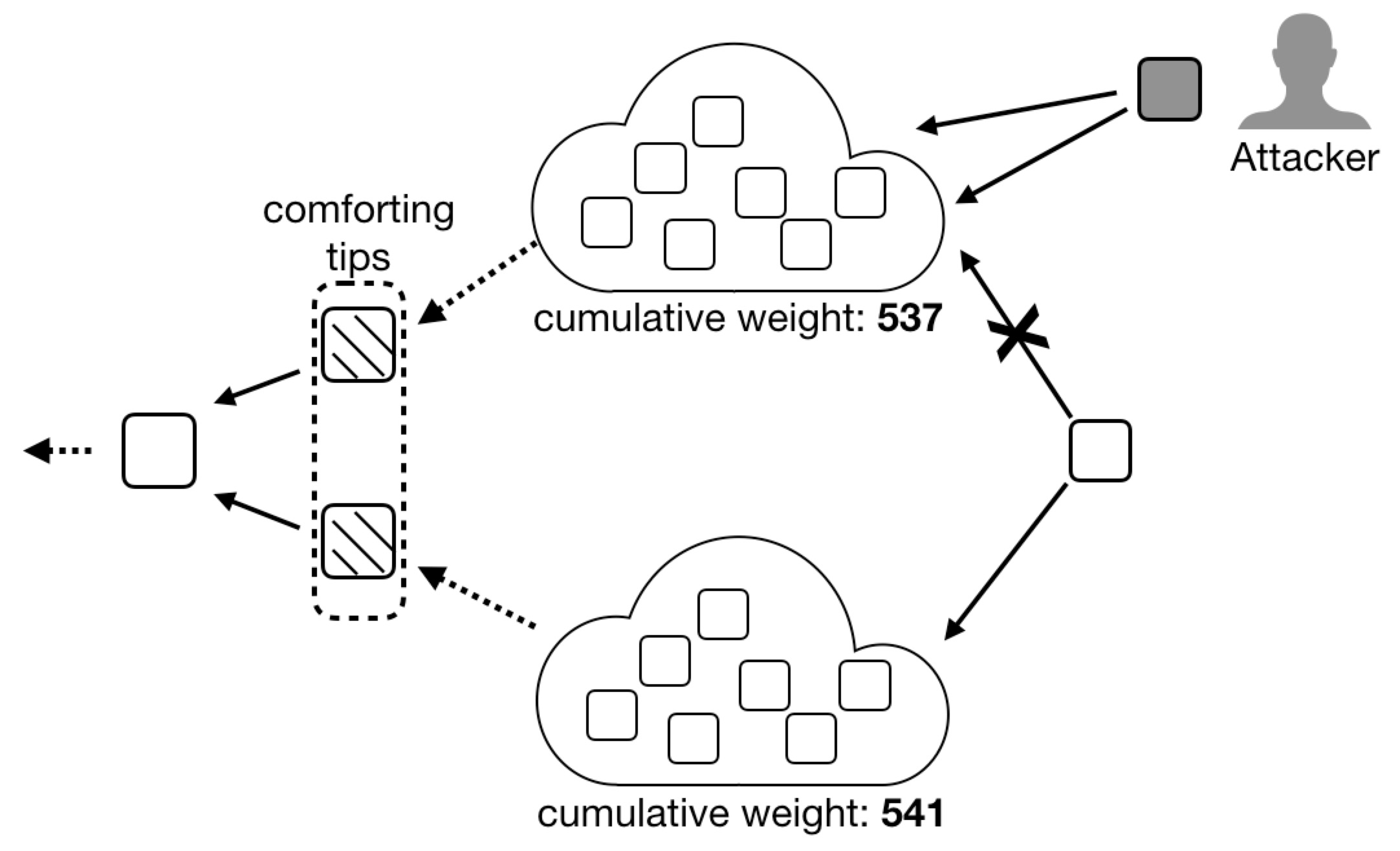}
\caption{Illustration of a splitting attack scenario.}
\label{splitting}
\end{figure}

In a splitting attack, the attacker splits  the tangle   into two branches so that s/he can launch double spending in these two different branches. 
When the total cumulative weight of one branch goes down compared to the other branch, the attacker generates several meaningless transactions to keep the balance between these two branches. 
Figure \ref{splitting} illustrates a splitting attack. 
The attacker first generates two conflicting transactions attached to the same previous transactions. 
Any honest user who wants to verify both of them has to chose only one of them during the tips selection. 
Two independent branches therefore are generated automatically by the users of IOTA. 
As these two branches independently belong to two different verification paths, a pair of double spending transactions appended to one and another will both be considered as correct transactions, if these two branches keep growing evenly. 
As the probability of having the same total cumulative weight for these two branches is very low, without additional intervention, more and more honest transactions will chose the branche with higher total cumulative weight according to weighted random walk tip selection algorithm. 
However, the attacker will keep the balance between these two branches by submitting meaningless transactions to the branch having the smaller total cumulative weight.

To avoid splitting attacks, a high $\alpha$ value is needed for weighted random walk tips selection. So that it will be very hard for an attacker to keep the balance between two branches. 
However, the higher $\alpha$ is, more tips could be \textit{left behind} according to the Weighted Random Walk algorithm given in Section \ref{sec:Tip-Selection-Algorithm}.



\begin{figure}
\centering
\includegraphics[width=0.45\textwidth]{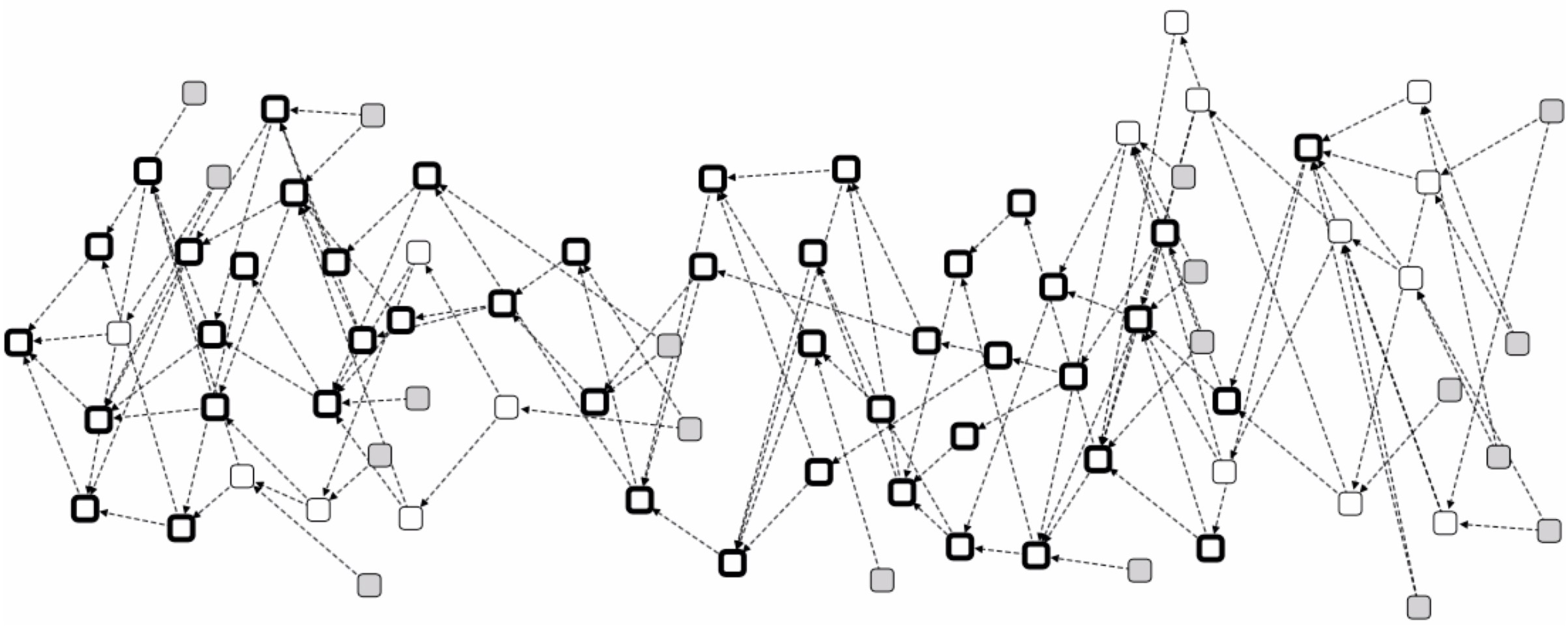}
\caption{An example of tangle with a relatively high $\alpha = 0.7$. As can be seen, many tips are left behind.}
\label{left}
\end{figure}

Figure \ref{left} shows a tangle with a relative high $\alpha$. Grey vertexes are tips, we can consider the tips in the left side as tips that are left behind, and the tips in the right side as tips that are not left behind (for the moment).

Using the weighted random walk algorithm with relatively high $\alpha$, the tangle is resilient  to splitting attacks but converges to an important rate of left behind tips. This means that the honest tips who have been left behind cannot get any chance to be verified except for the tips that are re-submitted. A similar conclusion is also presented in \cite{kusmierz2018probability}.

We define the level of \emph{confidence fairness} at a time $t$, $CF_t$, for a tips selection algorithm in a given IOTA-tangle network configuration. 
That is $1$ minus the proportion of none left-behind transactions to all transaction at time instant $t$.


\begin{figure}[h]
\centering
\includegraphics[width=0.45\textwidth]{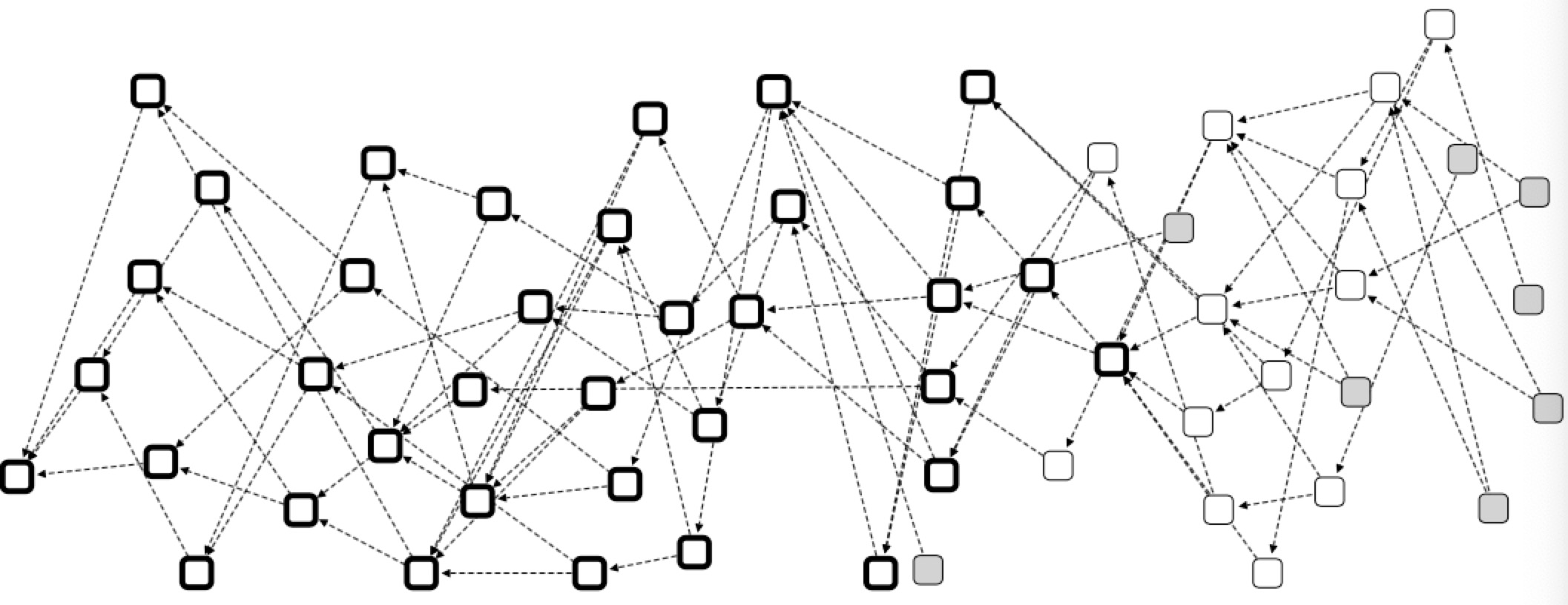}
\caption{An example of tangle with relative low $\alpha = 0.1$: very few tips are left behind.}
\label{noleft}
\end{figure}

The weighted random walker tips selection with high $\alpha$ therefore, according to our analysis, has a lower confidence fairness, compared with low $\alpha$ case shown in Figure \ref{noleft}.
 
\section{G-IOTA} 
\label{sec:G-IOTA}

According to the analysis given in the previous section, \ref{sec:Drawbacks-of-IOTA}, recent tips selection weighted random walk used in IOTA-tangle gives a low \textit{fairness} when choosing a relative high $\alpha$ to resist splitting attack. 
For this reason, we propose G-IOTA that allows honest transactions in a tangle to increase their confidences evenly and quickly to meet the requirement of high \textit{confidence fairness}. 
Implicitly, this way, no honest tips and transactions will be left behind during the growth of the tangle. 
Transactions with low confidence for a relatively long time can  be considered as fake transactions. 
Therefore, fake transactions can be detected easily which easy to implement a  punishment mechanism in G-IOTA.

\subsection{Transaction confidence and Tips selection algorithm}
\label{relation}

Our high confidence fairness tips selection algorithm aims at increasing the confidence of all honest transactions. 
In the following we will discuss the relation between Transaction confidence and Tips selection algorithm. 

From the analysis proposed in \cite{popov2016tangle}, we conclude four \emph{important facts} that could help understanding the relation between confidence and tips selection.

\begin{def5}
\label{lemma1}
The confidence of the majority of the honest transactions increases with the evolution  of  the tangle.
\end{def5}

Note that  each transaction approves two other transactions. It is clear that with the increase of the depth of a tangle, a given transaction has an increased probability to be  verified directly or indirectly by the recent tips. 

\begin{def5}
\label{lemma2}
Transactions will hold their confidences once it reaches 100\% with high probability.
\end{def5}

That is, when the confidence of a transaction $s$ goes to 100\%,  all the tips in $A_s$ selected by $n$ runs of tips selection algorithm verify $s$ directly or indirectly. 
Then, when new tips come into the tangle they will chose  with high probability tips to verify only among tips in $A_s$. 

\begin{def5}
\label{lemma3}
The confidence of a transaction $s$ is at least equal to the maximum confidence of a transaction among all the transaction verifying $s$ directly or indirectly.
\begin{align} 
\label{fff2}
C_s >= \max \{... C_q ...\} \quad \forall q \rightarrow s
\end{align}
\end{def5}
where the symbols $q \rightarrow s$ means transaction $q$ approves $s$ directly or indirectly.

That is,
consider  $p$ verifies transaction $s$ directly or indirectly. 
If $C_p = p\%$, that means that by launching a lot of tips selection algorithm, $p\%$ tips verify $p$. 
As $p$ verifies $s$, that means selected tips also verify $s$. 
The confidence of $s$ therefore is at least equal to the maximum confidence of a transaction among all the transactions verifying $s$.

\begin{def5}
\label{lemma4}
If a none-tip transaction $s$ has been left behind, transactions $q \quad \forall q \rightarrow s$ will also be left behind. 
And there must be at least one tip approving $s$ having a low probability to be chosen by the tips selection algorithm, which means these tips will be left behind with high probability. 

If one tip has been left behind, there could be some none-tip transaction approved by it that has been left behind.
\end{def5}

That is, according to Note \ref{lemma3}, all transactions verifying the left-behind transaction $s$ could not have higher confidence than $s$ has, they will therefore be left behind in the future with high probability. The reason that a non-tip transaction is left behind is that this transaction cannot get new verifications from incoming tips. That means that the current tips approving $s$ have very low probabilities to be selected by the tip selection algorithm.

If a tip has been left behind, it could lead to some left-behind transaction, but not for sure because non-tips transactions approved by it could have other tips that could be chosen by the tips selection algorithm.



To increase the confidence fairness and to reduce de number of left-behind transactions, there are three intuitive ways, according to our analysis above:

\begin{enumerate}
\item Increase the number of non left-behind tips that verify the left-behind transactions directly or indirectly.
\item Increase the probabilities that tips verifying left-behind transactions directly or indirectly can be chosen by the tips selection algorithm.
\item Increase the confidence of $q$, $C_q \quad \forall q \rightarrow$ the left-behind transactions.
\end{enumerate}

\subsection{Confidence fairness aware tips selection algorithm}

According to the analysis detailed in Section \ref{relation}, we propose a new confidence fairness aware tip selection algorithm G-IOTA. 
This new algorithm is based on the weighted random walk tips selection. In addition, we integrate a \emph{Left-behind Tips Protection} mechanism that allows tips who have been left behind regain the opportunity to be approved by incoming tips, which further decreases the transactions left behind. 

The G-IOTA tip selection algorithm is as follows: 

\begin{enumerate}
\item Launch the weighted random walk tip selection algorithm.
\item Chose the two tips selected by the tips selection algorithm. Note them as $tips_1$ and $tips_2$.
\item Search in the tangle tips left behind. 
\item Chose one left-behind tip whose verification path contains the smallest average confidence (among all non-confirmed transactions) as the candidate of $tips_3$.
\item Approve $tips_3$, if it is conflicting with verification paths of $tips_1$ and $tips_2$, then postpone the candidate to the one who has the second smallest average confidence, until we find one candidate for $tips_3$ that does not conflict with $tips_1$ and $tips_2$. $tips_3$ therefore is the chosen left-behind tips.
\item Chose $tips_1$, $tips_2$ and $tips_3$ as three selected tips, if $tips_3$ exists; chose only $tips_1$ and $tips_2$, otherwise.


\end{enumerate}





Note that our new tips selection algorithm allows each new transaction approve additionally a left-behind tip if there is any. 
This mechanism allows left-behind tips to be approved by new incoming transactions. 
The reason is simple: when a new transaction $p$ arrives and choses two first tips, $tips_1$ and $tips_2$, following the tips selection algorithm, this new transaction will have a relative high probability to be approved in the future according to Note \ref{lemma1}. If $p$ adopts the left-behind tips protection and approve one more left-behind transaction $s$ then $s$ will also get the chance to be approved in the future, because the confidence of $s$, $C_s$ is at least equal to $C_p$, according to the Note \ref{lemma3}. 

\begin{figure}
\centering
\includegraphics[width=0.45\textwidth]{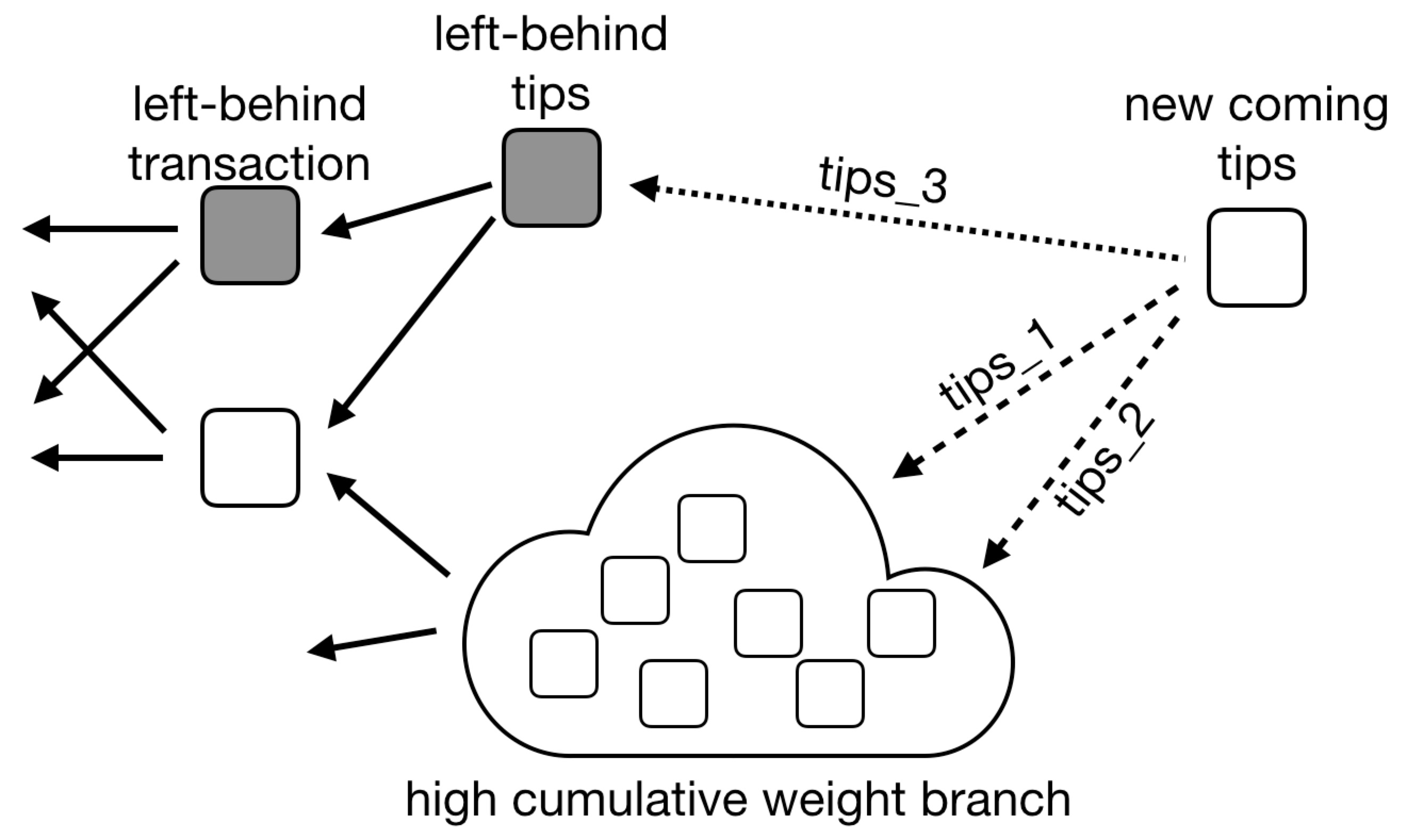}
\caption{Tips selection with left-behind tips protection}
\label{protection}
\end{figure}

Figure \ref{protection} shows how the left-behind tips protection gives the opportunity to the left-behind tips and transactions to increase  their confidence. 
At a given instant, a partial view of a tangle is as follows. 
Ignore the new incoming tips, the two grey vertexes are left-behind transaction and tips. 
The reason for that is that the branches they belong have significantly smaller cumulative weights compared to the other branches. 
The probability that grey tips can be chosen by tips selection algorithm is therefore significantly smaller than for the other tips in another branch. 
Then a new transaction (will be a new tip) comes. 
Very likely, according to weighted random walk tips selection algorithm, it choses $tips_1$ and $tips_2$ among the tips in the lower branch in the figure. 
As there is a left-behind tip present in the tangle, the incoming tips need to approve it respecting the new tips selection algorithm description. 
In this way, the incoming tips verify a left-behind tip, so that this left-behind tips regain at least the confidence of the incoming tips, and so does the left-behind transaction.
 As the new tips verify $tips_1$ and $tips_2$ in the branch with a higher cumulative weight, these new tips therefore can continuously increase their confidence according to Note \ref{lemma1}.

By using the G-IOTA tips selection, all honest tips will be \textit{guaranteed} to receive one first approve in a fixed difference of depth, $D_S$, the threshold in the definition of left-behind tips. The G-IOTA tips selection therefore holds a very high confidence fairness since the left-behind transactions regain their confidence with the evaluation of tangle.

\section{Discussions}
\label{sec:Discussion}

In this section, we discuss and analyze the strengths and the weaknesses of G-IOTA.

\subsection{Incentives and Punishments}
\label{encourage}
We show that our new tip selection algorithm has a good confidence fairness. 
However, why would a user do additional verification to approve the third tips to protect left-behind tips? 
A reasonable incentive is that user helping other transactions to regain the opportunity being approved will also receive help from others if the honest users are the majority.

Also, in the process of choosing only one tip to approve from two conflicting verification paths during the choosing of $tip_1$ and $tips_2$, we can chose the one whose verification path contains more transaction carrying out the left-behind tips protection mechanism as a additional reward. 
Because verification paths with more transaction carrying out the left-behind tips protection means more honest user choosing this path to attach their transaction. 
Note that by helping left-behind tips, the whole confidence in the tangle of honest transactions will increase fast. 
The conflicting transactions, on the other hand, will be left-behind fast, because honest users will not verify them. 
We can therefore recognize and locate these conflicting transactions if they always hold low confidence for an unusual long time duration ($T >> S_d$). 
This big different confidence between honest transactions and fake transactions can be used further as punishment mechanism.

\subsection{Mutual Supervision Mechanism}
\label{speculator}

What if a speculative user always verify a third tip, a non-left-behind tip, to pretend it works hard for helping the others? Or what if a lazy user only chooses non-left-behind transactions to verify and pretends his/her new transaction has become a left-behind tips. This lazy user then waits for others to save that transaction? 

Here we propose a \emph{Mutual Supervision Mechanism} that allows the user in the G-IOTA system detect speculative or lazy behaviour. 
The idea is that, when a user receives a new submitted transaction sent to him, it can look into his/her local tangle, if once the tips selected, $tips_1$, $tips_2$ or $tips_3$ is not a left-behind tips for a long time, meaning that this transaction has at least one path with length $L$ to one tip. 
The receiver can block this transaction or announce the sender that, s/he is cheating. By the supervision of all honest users in the G-IOTA system, the speculative and lazy users cannot profit from their behaviours.

Note that, $L$ depends on the network latency. In a high-latency network environment, different nodes could have a different view between their local views and the real view of world tangle. 
That leads an honest user think that a transaction is still left-behind in the tangle, even it has already been protected by others in the real work tangle, we therefore can use a relatively high $L$ to tolerant this case. 
However, in a low-latency network environment, every user normally receives new transactions and flash their local tangles very fast. We say that the local view and the real view of tangle has a very small difference. 
In this case, $L$ could be relatively small because honest users will not be affected by the network latency. 
More possible impacts of the network latency to the G-IOTA network is also an interesting future work.


\subsection{Attack Analysis}
The G-IOTA tip selection algorithm follows basically the same idea as Weighted Random Walk except for the left-behind tips protection mechanism. This new algorithm still takes two recent tips as the two parents following the same way of Weighted Random Walk. 
The new tip selection therefore tolerates all the attacks mentioned in \cite{popov2016tangle}. 
In addition, our left-tips protection mechanism allows higher $\alpha$ than Weighted Random Walk of the original IOTA protocol. Therefore, it has better resistance to splitting attacks.

\section{Conclusion and Future Work}
\label{sec:Conclusion-and-Future-Work}

In this work, we proposed G-IOTA, a new tips selection mechanisms that combines a confidence fairness aware tips selection algorithm and a mutual supervision mechanism. The new tips selection algorithm, chooses not only two tips as the classical IOTA but one more tip, which is a left-behind tip in the tangle. 
That allows increasing the fairness in terms of transaction confidence for all honest transactions in tangle and guarantees the first approval for all honest tips.
Integrated incentive and mutual supervision mechanisms guide users to abide by the proposed algorithm. In addition, the new tips selection algorithm  tolerates all the  attacks tolerated by IOTA-tangle, including the splitting attack.

 
As future work we intend to extend the analysis proposed in \cite{popov2017equilibria}  in order to prove that G-IOTA tips selection is a \emph{Nash Equilibria}. 

\bibliographystyle{plain}
\bibliography{sample-dmtcs}

\end{document}